\documentclass[prl,amsmath,amssymb,reprint,nofootinbib]{revtex4-1}

\usepackage{graphicx}
\usepackage{color}
\usepackage{amsmath, slashed}

\begin{document}
\title{$R_{D^{(*)}}$ anomaly: A possible hint for natural supersymmetry with $R$-parity violation} 

\author{Wolfgang Altmannshofer$^1$, P. S. Bhupal Dev$^{2}$, Amarjit Soni$^3$}
\affiliation{$^1$Department of Physics, University of Cincinnati, Cincinnati, OH 45221, USA}
\affiliation{$^2$Department of Physics and McDonnell Center for the Space Sciences,
Washington University, St. Louis, MO 63130, USA}
\affiliation{$^3$Physics Department, Brookhaven National Laboratory, Upton, NY 11973, USA}

\begin{abstract}
Recently, several $B$-physics experiments have reported an appreciable 
deviation from the Standard Model (SM) in the tree-level observables $R_{D^{(*)}}$; the combined weighted average now stands at $\approx 4 \sigma$.  We first show the anomaly necessarily implies model-independent collider signals of the form $pp \to b \tau \nu$ that should be expediously searched for at ATLAS/CMS as a complementary test of the anomaly. Next we suggest a possible interconnection of the anomaly with the radiative stability of the Standard Model Higgs boson and point to a minimal effective supersymmetric scenario with $R$-parity violation as the underlying cause. We also comment on the  possibility of simultaneously explaining the recently reported $R_{K^{(*)}}$ anomaly in this setup.
\end{abstract}

\maketitle

\section{Introduction}

In the past few years, several $B$-physics experiments, such as BaBar, LHCb and to a lesser degree Belle, seem to find an extremely interesting and surprising anomaly in simple semi-leptonic decays of $B$-mesons~\cite{Hirose:2017vbz,talk2}.
The ratios, $R_{D^{(*)}}$ of the branching ratios
\begin{equation}
 R_D = \frac{\mathcal{B}(B \to D \tau \nu)}{\mathcal{B}(B \to D \ell \nu)} \, ,\quad  R_{D^*} = \frac{\mathcal{B}(B \to D^* \tau \nu)}{\mathcal{B}(B \to D^* \ell \nu)} ~,
\end{equation}
where $\ell = e,\mu$ for BaBar and Belle, while $\ell = \mu$ for LHCb,
indicate appreciable deviations from the Standard Model (SM) expectations. These ratios are independent of CKM angles and have strongly reduced hadronic uncertainties.
This simple comparison of $B$-decays to $\tau$ with decays to $\mu$ or $e$ has long been suggested~\cite{Kiers:1997zt, Chen:2006nua, Nierste:2008qe, Kamenik:2008tj}
as sensitive probe of new physics (NP) beyond the SM.
The experimental world averages of BaBar~\cite{Lees:2012xj,Lees:2013uzd}, Belle~\cite{Huschle:2015rga,Sato:2016svk,Abdesselam:2016xqt, Hirose:2016wfn, Hirose:2017dxl}, and LHCb~\cite{Aaij:2015yra,new} from the Heavy Flavor Averaging Group~\cite{Amhis:2016xyh} read
\begin{align}
R_D^\text{exp} & = 0.403 \pm 0.040 \pm 0.024 \, ,\nonumber \\
R_{D^*}^\text{exp} & = 0.310 \pm 0.015 \pm 0.008 \, ,
\label{hfag}
\end{align}
with a mild negative correlation of $-0.23$. One recent SM prediction for these ratios reads~\cite{Bernlochner:2017jka}
\begin{equation} \label{RDRDstarSM}
  R_D^\text{SM} = 0.299 \pm 0.003 ~,~~  R_{D^*}^\text{SM} = 0.257 \pm 0.003~.  
\end{equation}
with an error correlation of $+0.44$; see also~\cite{Fajfer:2012vx,Na:2015kha,Lattice:2015rga,Bigi:2016mdz} for earlier predictions.
There is another recent~\cite{Bigi:2017jbd} phenomenological study of the SM prediction, which finds
$R_{D^*}^\text{SM} = 0.262 \pm 0.010$.

Lattice computations of the $B \to D$
form factors~\cite{Na:2015kha,Lattice:2015rga,Aoki:2016frl} allow a precise determination of $R_D$; 
{\it e.g.} Ref.~\cite{Na:2015kha} get $R_D = 0.299 \pm 0.011$. No complete lattice results for the $B \to D^*$ form factors are so far available; an existing study~\cite{Bailey:2014tva} only tackles the form factors at the end-point. 
On the lattice, the $B \to D^*$ transition is significantly more involved because it has four form-factors rather than two for $B \to D$. Also, the $D^*$ undergoes strong (and electromagnetic) decay further complicating a study. For these reasons, from the lattice perspective, the $B \to D^*$ uncertainties should be appreciably bigger compared to the  $B \to D$ case. 
Nevertheless, using heavy quark effective theory and measured $B \to D^{(*)} \ell \nu$ decay distributions
seem to lead to controlled predictions also for $R_{D^*}$~\cite{Fajfer:2012vx,Bernlochner:2017jka,Bigi:2017jbd}.
Bearing all this in mind, in our study we will take the SM prediction to be, ($R_D^\text{SM}, R_{D^*}^\text{SM}) = (0.299 \pm 0.011, 0.260 \pm 0.010)$.

Note that for now the experimental uncertainties are considerably larger than the theory uncertainties on the SM predictions.
The combined difference of measurements and SM predictions is
stated to be 
$\approx 4.1\sigma$~\cite{new}.

The most significant enhancements of $R_{D^{(*)}}$ are seen in the BaBar analyses~\cite{Lees:2012xj,Lees:2013uzd}. The other results also show enhanced $R_{D^{(*)}}$, but are less significant, especially from Belle~\cite{Huschle:2015rga,Sato:2016svk,Abdesselam:2016xqt, Hirose:2016wfn, Hirose:2017dxl}. 
In fact, recent Belle~\cite{ Hirose:2017vbz, Hirose:2017dxl} and LHCb~\cite{new} results obtained by using $\tau$ decays to hadron(s) $+ \nu$ tend to deviate considerably less from the SM prediction. This may be especially significant as such hadronic decays of tau entail only one neutrino rather than two into the leptonic modes and so may be cleaner systematically. Be that as it
may, 
these experimental results seem to suggest lepton flavor universality violation (LFUV) and therefore have  received much attention as a possible hint of NP; 
see e.g.~\cite{Fajfer:2012jt, Datta:2012qk, Bailey:2012jg, Tanaka:2012nw, Biancofiore:2013ki, Alonso:2015sja, Greljo:2015mma, Calibbi:2015kma, Freytsis:2015qca, Bhattacharya:2015ida, Alonso:2016gym, Nandi:2016wlp, Ivanov:2016qtw, Ligeti:2016npd, Bardhan:2016uhr, Bhattacharya:2016zcw, Alonso:2016oyd, Celis:2016azn, Bordone:2017anc} for model-independent studies and~\cite{Bauer:2015knc, Hati:2015awg, Fajfer:2015ycq, Barbieri:2015yvd, Cline:2015lqp, Zhu:2016xdg, Boucenna:2016wpr, Das:2016vkr, Li:2016vvp, Boucenna:2016qad, Deshpand:2016cpw, Becirevic:2016yqi, Sahoo:2016pet, Hiller:2016kry, Bhattacharya:2016mcc, Wang:2016ggf, Popov:2016fzr, Barbieri:2016las, Wei:2017ago, Cvetic:2017gkt, Ko:2017lzd, Chen:2017hir, Chen:2017eby, Megias:2017ove, Crivellin:2017zlb} for recent discussions of specific NP models.
While the deviations may well be due to lack of statistics and/or systematic or theoretical issues that need~\cite {Nandi:2016wlp} further understanding, we take them at face value and explore the potentially exciting theoretical consequences.
In this context, it is useful to  remind ourselves that the common folklore does {\it not} expect new phenomena to first show up in such simple tree-level decays. The widely held belief has been that in flavor-changing neutral current (FCNC) transitions, wherein one is able to access very short-distance physics because of the uncertainty principle, effects of NP are expected to show up first. 
Thus, one is led to ponder what is so special about these semi-leptonic decays. In the following, we take the hint of these tree-level decays into account in motivating a NP model. 

We address two important issues that these interesting findings indicate. 
First, it is clearly of paramount importance that more experimental information be accumulated to solidify these findings as soon as possible. While the ongoing $B$-experiments at LHCb and the forthcoming Belle II will clearly be addressing these, here we propose a completely model-independent class of searches that the LHC experiments at the high-energy frontier, i.e.~ATLAS and CMS, can do to enlighten us on these anomalies.
The underlying reaction in the $B$-experiments that indicates the anomalous behavior is the weak decay, $b \to c \tau \nu$, which is CKM-suppressed in the SM.  
Therefore, in complete generality ({\it i.e.} without recourse to any specific NP model), we also expect anomalous behavior in the basic reaction,  
$gc  \to b\tau\nu$, where $g$ is the gluon, in $pp$ collisions.
This robust connection empowers collider experiments to complement the aforementioned $B$-experimental studies and search {\it directly} for the signals of NP.

Second, since these decays involve the $\tau$ and $b$, both members of the third fermion family, it might have a deeper meaning. In fact, a pressing problem in Particle Physics is presented by the third family, namely, the top quark which makes the dominant contribution to the self-energy of the Higgs boson and consequently is intimately related to the Higgs radiative stability and the naturalness problem of the SM.  A famous candidate for addressing the naturalness problem is, of course,  supersymmetry (SUSY).
However, given the null results in direct SUSY searches at the LHC so far~\cite{Autermann:2016les}, SUSY solutions to naturalness have become less elegant. In a drive for simplicity and minimality we assume that only the third generation is {\it effectively} supersymmetric at the low-scale.
An important consequence of this minimal construction is that constraints on $R$-parity violating (RPV) couplings from proton decay get relaxed.

An extremely attractive feature of SUSY is gauge coupling unification. We explicitly show that the minimal version of effective SUSY that we are invoking does indeed retain this unique attribute. The added advantage of this scenario is that it provides a simple solution to the $R_{D^{(*)}}$ puzzle alluded to above, and also alleviates the naturalness problem of the SM and the flavor and CP problems of the minimal supersymmetric SM (MSSM).

\section{Model-independent Collider Analysis}

There is an important and unique collider signal directly implied by the $R_{D^{(*)}}$ anomaly. At the parton level, the reaction $b\to c\tau {\nu}$ necessarily implies by crossing symmetry that the process $gc\to b\tau{\nu}$ should also take place at the LHC.\footnote{In some models, $b\to c\tau{\nu}$ is also related by $SU(2)$ symmetry to the LHC process  $b \bar b \to \tau^+\tau^-$~\cite{Faroughy:2016osc}.}

\begin{figure*}[ht!]
\includegraphics[width=0.46\textwidth]{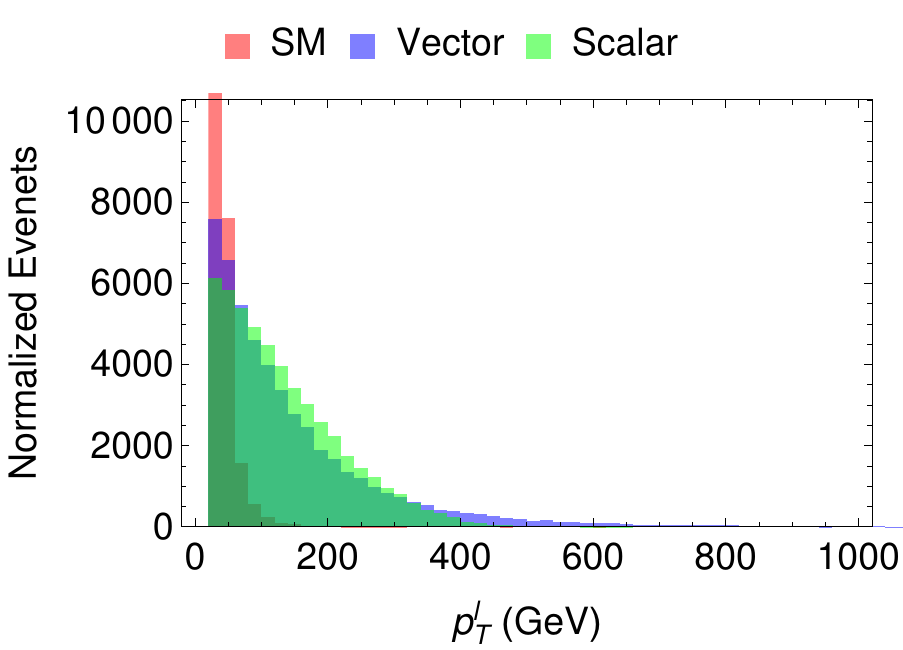}~~~
\includegraphics[width=0.46\textwidth]{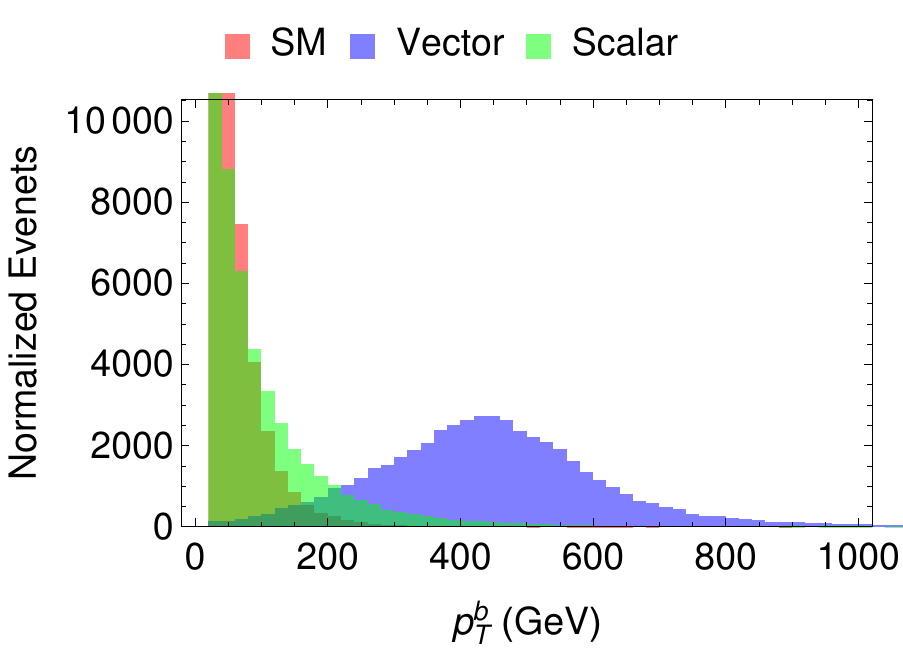}\\[12pt]
\includegraphics[width=0.46\textwidth]{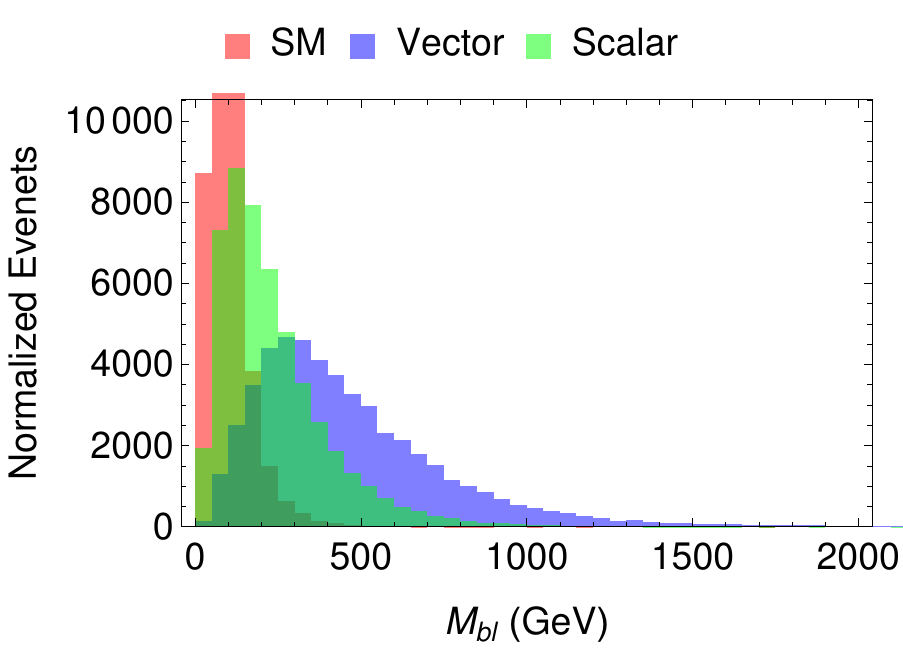}~~~
\includegraphics[width=0.46\textwidth]{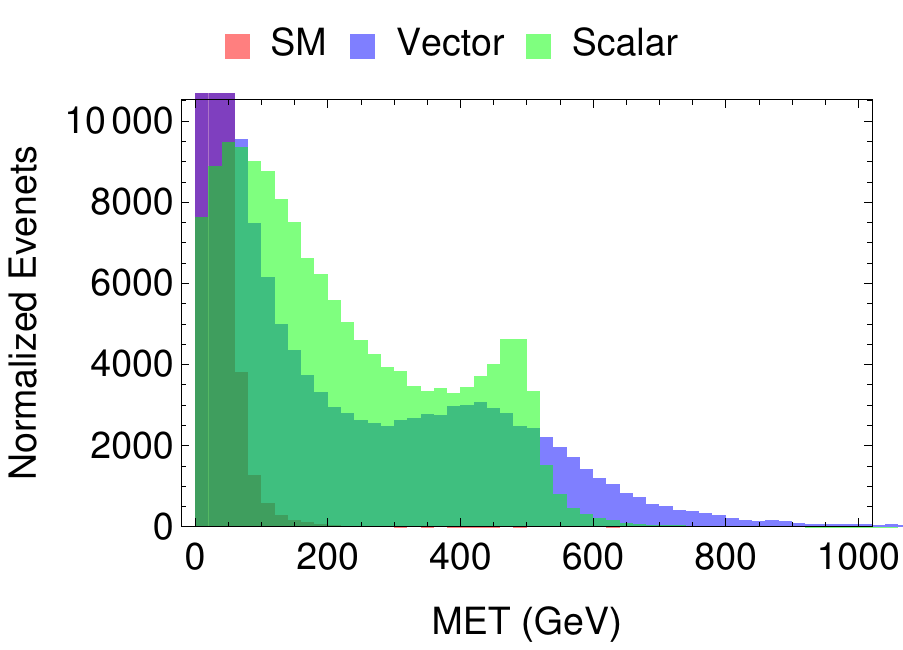}
\caption{Normalized kinematic distributions for the $pp\to b\tau\nu \to b\ell +\slashed{E}_T$ signal and background.}
\label{fig:dist}
\end{figure*}

At low energies, the effective 4-fermion Lagrangian for $b\to c\tau {\nu}$ in the SM is given by 
\begin{align}
-{\cal L}_{\rm eff} \ = \ \frac{4G_F V_{cb}}{\sqrt{2}}\left(\bar{c}\gamma_\mu P_L b\right)\left(\bar{\tau} \gamma^\mu P_L \nu_{\tau}\right)+{\rm H.c.} \, ,
\label{eq:4fermi}
\end{align}
where $G_F$ is the Fermi constant, $P_{L}\equiv (1-\gamma_5)/2$ is the left-chiral projection operator, and $V_{cb}\approx 0.04$.
In the SM, the $gc\to b\tau{\nu}$ rate is also CKM-suppressed by $|V_{cb}|^2$, while in a generic NP scenario, this need not be the case, which might make it possible for the NP signal to be observable above the SM background at the LHC. 
The same final state also arises in $gu \to b \tau {\nu}$, which is suppressed by $|V_{ub}|^2$, but enhanced by the $u$-quark density in the proton.

In a realistic hadron collider environment, one must consider other potential backgrounds, such as 
\begin{enumerate}
\item [(i)] the charged-current process $pp \to j W \to j \tau {\nu}$, where $j$ stands for a light quark (or a gluon) jet misidentified as a $b$-quark jet, 
\item [(ii)] $pp\to W\to \tau {\nu}$, with an initial state radiation of gluon which is then split into $b\bar{b}$ and one of the $b$-quarks is lost, 
\item [(iii)] the single-top production $pp\to tj\to b\tau{\nu}j$ and $pp\to tW\to b\tau {\nu}jj$, where the jet(s) are lost, and 
\item [(iv)] $pp\to b\bar{b} j$, where one $b$-quark is misidentified as a $\tau$-lepton and the light jet is lost (i.e. misidentified as missing transverse energy). 
\end{enumerate}

The $b$-jet misidentification rate at the LHC typically varies between 1-10\%, depending on the $b$-tagging efficiency~\cite{Chatrchyan:2012jua}. We will assume a $b$-tagging efficiency of 70\%, for which the probability of light-parton misidentification as a $b$-quark is about 1.5\%. Similarly, the probability that one or more jets (including the $b$-jets) are lost in the detector and that a $b$-quark is misidentified as a $\tau$-lepton are both assumed to be at the percent level~\cite{Aad:2009wy, ECFA}.
Combining all of this and imposing basic trigger cuts on the lepton and jet transverse momenta $p_T^{j,b,\ell} >20$ GeV, missing transverse energy $\slashed{E}_T>20$ GeV, pseudo-rapidity $|\eta^{j,b,\ell}|<2.5$ and 
isolation cuts $\Delta R^{\ell j, \ell b, jb}>0.4$ on simulated events obtained using the {\tt MadGraph5} event generator~\cite{Alwall:2014hca} (with the {\tt sm-lepton\_masses} and {\tt taudecay\_UFO} models to properly handle the $\tau$ decays), we find the cross section for the total SM background at $\sqrt s=13$ TeV LHC to be $\sigma_\text{SM}(pp\to b\tau{\nu}\to b\ell+\slashed{E}_T) \simeq 47$~pb (here we have considered only the 
leptonic decay of $\tau$ with $\ell=e,\mu$)\footnote{We thank Brian Shuve for pointing out an earlier error in our cross section estimate, which was caused due to the default value of zero $\tau$-width in {\tt MadGraph5}.}, where the dominant contributions come from the $pp\to Wj$ and $pp\to b\bar{b}j$ channels.

As for the NP contribution, we consider the following dimension-6 four-fermion operators~\cite{Freytsis:2015qca}: 
\begin{align}
{\cal O}_{V_{R,L}} \ & = \ \left(\bar{c}\gamma^\mu P_{R,L}b\right)\left(\bar{\tau}\gamma_\mu P_L\nu\right)\, \label{eq:vector} \\
{\cal O}_{S_{R,L}} \ & = \ \left(\bar{c}P_{R,L}b\right)\left(\bar{\tau}P_L\nu\right)\, . \label{eq:scalar}
\end{align}
The amplitudes for the collider process $gc\to b\tau{\nu}$ are suppressed by $g_{\rm NP}/\Lambda^2$, where $g_{\rm NP}$ denotes the effective NP coupling in the contact interaction and $\Lambda$ is the NP scale. 
For a typical choice $g_{\rm NP}/\Lambda^2=(1~{\rm TeV})^{-2}$, we obtain a signal cross section for $pp\to b\tau{\nu}\to b\ell+\slashed{E}_T$ of $\sigma_V\simeq 1.1$~pb for the vector case and $\sigma_S\simeq 1.8$~pb for the scalar case, both at $\sqrt s=13$ TeV LHC. These cross section estimates imply that even {\it without} using any specialized selection cuts to optimize the signal-to-background ratio, the NP signals associated with the $R_{D^{(*)}}$ anomaly may be directly probed at $3\sigma$ confidence level for mediator masses up to around 2.4 (2.6) TeV in the vector (scalar) operator case with ${\cal O}(1)$ couplings at $\sqrt s=13$ TeV LHC with an integrated luminosity of 300 fb$^{-1}$. 
 
The signal-to-background ratio can be improved in various ways. For instance, simple kinematic distributions, such as the transverse momentum of the outgoing $b$-quark (or of the final lepton) and the invariant mass of the $b$ quark and lepton system (see Fig.~\ref{fig:dist}), can be used to distinguish the NP signals from each other and from the SM background for different NP operators. Furthermore, imposing stringent cuts like $p_T^b>100$ GeV and $M_{b\ell}>100$ GeV could drastically reduce the SM background, without significantly affecting the signal (see Fig.~\ref{fig:dist}), especially in the vector case, potentially enhancing the LHC sensitivity to even higher mediator masses. Similarly, increasing the $\slashed{E}_T$ cut to 100 GeV will significantly reduce the SM background, including the mis-measured dijets, without much signal loss, as can be seen from Fig.~\ref{fig:dist}. For illustration, we show in Tab.~\ref{tab:cut} the individual cut  efficiencies of the signal and background 
for three representative values of the kinematic cuts for the four kinematic observables considered in Fig.~\ref{fig:dist} (taken one at a time). A detailed cut optimization study with all these (and possibly more) variables taken together could be done, e.g.~using a multivariate analysis or a Boosted Decision Tree algorithm, which is probably best dealt with by experimentalists possessing the relevant expertise.

\begin{table}[!t]
\begin{center}
\begin{tabular}{c|c|c|c|c}\hline\hline
 & Cut   & \multicolumn{3}{c}{Efficiency} \\ \cline{3-5}
Observable & value & SM & Signal  & Signal \\ 
& (GeV) & background & (Vector case) &  (Scalar case) \\  
\hline 
& 100 & 0.01 & 0.52 & 0.56 \\
$p_T^\ell$ & 50 & 0.10 & 0.78 & 0.82 \\
& 30 & 0.44 & 0.92 & 0.94 \\ \hline
& 100 & 0.13 & 0.99 & 0.33 \\
$p_T^b$ & 50 & 0.47 & 1.00 & 0.62 \\
& 30 & 0.75 & 1.00 & 0.84 \\ \hline
& 100 & 0.18 & 0.96 & 0.76 \\
$M_{b\ell}$ & 50 & 0.63 & 0.99 & 0.94 \\
& 30 & 0.88 & 1.00 & 0.98 \\ \hline
& 100 & 0.01 & 0.54 & 0.70 \\
$\slashed{E}_T$ & 50 & 0.09  & 0.70 & 0.86 \\
& 30 & 0.29 & 0.79 & 0.92 \\ \hline
\hline\hline
\end{tabular}
\caption{Signal and background cut efficiencies for the kinematic variables shown in Fig.~\ref{fig:dist}.} 
\label{tab:cut}
\end{center}
\end{table}

The new collider signal $pp\to b\tau{\nu}$ proposed here would be a powerful model-independent 
check of the $R_{D^{(*)}}$ anomaly\footnote{Here we are assuming that the NP affects the modes involving taus. To be clear, a completely model independent crossing symmetry test of $R_{D(*)}$ requires comparison of the (differential) cross-section of $pp \to b \tau \nu_{\tau}$ to that of $pp \to b \ell \nu_\ell$ with $\ell =\mu, e$ via analogous ratios. In this case, for the relevant high energies, the lepton masses -- including the $\tau$ -- are negligible so the effective ratios should be unity in the SM.} and would imply a directly accessible mass range of the associated NP at the LHC. Further distinctions between the NP operators~\eqref{eq:vector}-\eqref{eq:scalar} could in principle be made using the tau polarization measurements, both in the LHC experiments~\cite{Aad:2012cia} and in $B$-physics experiments~\cite{Hirose:2016wfn,Alonso:2016gym,Ligeti:2016npd,Ivanov:2017mrj,Alonso:2017ktd}. But a detailed discussion of this, including a more realistic collider simulation with all detector smearing effects,
 is beyond the scope of this paper and might be studied elsewhere.

\section{Minimal SUSY with RPV}

As Higgs naturalness involves the third family fermions, we propose an economical setting, where only the third family is effectively supersymmetrized, with the corresponding sfermions and all gauginos and Higgsinos close to the TeV scale. The correction to the Higgs mass from the top-quark loop is canceled by the light stop contribution. 
The first two generation sfermions can be thought of being decoupled from the low-energy spectrum as in~\cite{Brust:2011tb,Papucci:2011wy}, and RPV arises naturally in this setup~\cite{Brust:2011tb}. 

\begin{figure}[t!]
\centering
\includegraphics[width=0.46\textwidth]{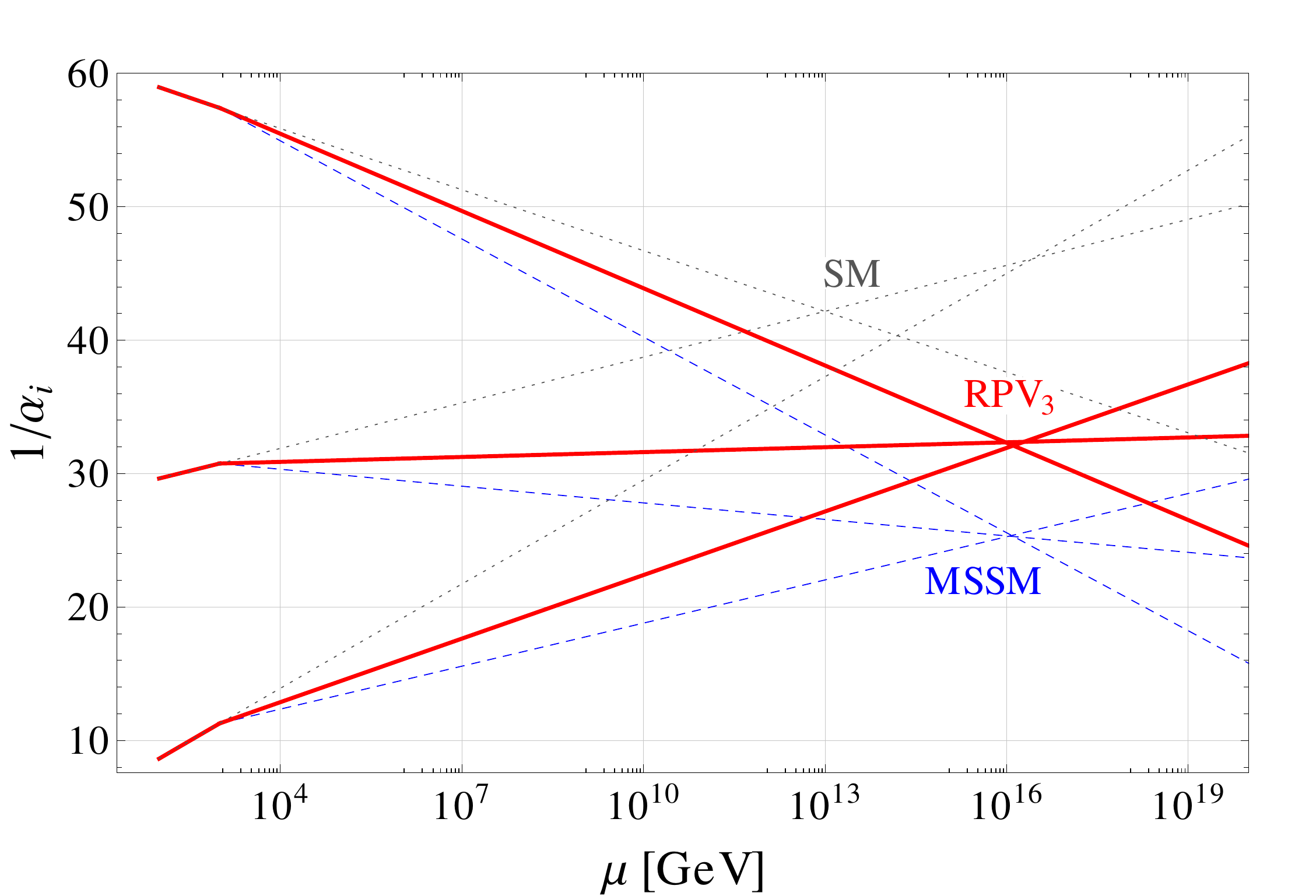}
\caption{RG evolution of the gauge couplings in the SM, MSSM and in our natural RPV SUSY scenario.}
\label{fig:uni}
\end{figure}

Despite the minimality of this setup, one of the key features of SUSY, namely, gauge coupling unification is still preserved, as shown in Fig.~\ref{fig:uni}. Here we show the renormalization group (RG) evolution of the inverse of the gauge coupling strengths $\alpha_i^{-1}=4\pi/g_i^2$ (with $i=1,2,3$ for the $SU(3)_c$, $SU(2)_L$ and $U(1)_Y$ gauge groups, where the hypercharge gauge coupling is in SU(5) normalization) 
in the SM (dotted) and the full MSSM with all SUSY partners at the TeV scale (dashed), and the RPV SUSY scenario with only third generation fermions supersymmetrized at the TeV scale (solid).\footnote{
The RG evolution in the SM and the MSSM is performed at the 2-loop level. In the RPV SUSY scenario we solve the RG equations consistently at 1-loop using the results from~\cite{Allanach:1999mh}. At higher loop level, the decoupled first and second generation squarks would require a refined analysis~\cite{Box:2007ss}, which is beyond the scope of our work, but our qualitative conclusions concerning gauge coupling unification are unaffected.
The impact of the RPV interactions on the running gauge couplings is small as long as the RPV couplings do not develop a Landau pole.}
Coupling unification occurs regardless of whether only one, two, or all fermion families are supersymmetrized at low scale, which only shifts the unified coupling value, but not the unification scale. This is valid, even in presence of RPV, as long as the gaugino and Higgsino sectors are not much heavier than the third family sfermions. 

In SUSY models, the Higgs mass parameter is related to the various sparticle masses. Requiring the absence of fine-tuned cancellations generically leads to upper bounds on sparticle masses. The Higgsino should not be heavier than a few hundred GeV, the stop mass should be well below a TeV and the gluino mass should not be far above a TeV~\cite{Barbieri:1987fn,Papucci:2011wy}. Bounds on other sparticle masses are considerably weaker. Nevertheless, also first and second generation sfermions are constrained from their two-loop contributions to the Higgs and $Z$ masses. A natural spectrum with less than $10\%$ tuning should have first 
and second generation squarks $\lesssim 10$~TeV~\cite{Buckley:2016tbs}. 
Allowing for $10^{-2}$ or even $10^{-4}$ tuning, bounds can be relaxed substantially. Thus, from the phenomenological point of view, we can decouple the first and second generation from the collider and flavor physics aspects being considered here.

\section{$B$-anomalies and constraints}

To explain the $R_{D^{(*)}}$ anomaly in the minimal RPV SUSY setup, we consider the $\lambda'$-couplings (see~\cite{Deshpande:2012rr,Zhu:2016xdg,Deshpand:2016cpw} for related studies):\footnote{We do not consider additional contribution to $R_{D^{(*)}}$ from charged Higgs exchange. Those contributions are small if the second Higgs doublet of the MSSM is heavy or if $\tan\beta$, the ratio of the two Higgs vacuum expectation values, is small. Even if we include this contribution, which involves the scalar operator~\eqref{eq:scalar} in the 4-fermion language, the model-independent collider signal discussed in the previous section provides a way to distinguish it from the squark contribution, which involves a vector operator~\eqref{eq:vector}, as explicitly shown in Eq.~\eqref{Leff}.}
\begin{align}
{\cal L}\ = \ & \lambda'_{ijk}\big[\tilde{\nu}_{iL}\bar{d}_{kR}d_{jL}+\tilde{d}_{jL}\bar{d}_{kR}\nu_{iL}+\tilde{d}^*_{kR}\bar{\nu}^c_{iL}d_{jL}\nonumber \\
-& \tilde{e}_{iL}\bar{d}_{kR}u_{jL}-\tilde{u}_{jL}\bar{d}_{kR}e_{iL}-\tilde{d}^*_{kR}\bar{e}^c_{iL}u_{jL}\big]+{\rm H.c.}
\end{align}
Working in the mass eigenbasis for the down-type quarks and assuming that sfermions are in their mass eigenstates, we obtain the following four-fermion operators at the tree-level after integrating out the sparticles (see also~\cite{Deshpand:2016cpw})
\begin{align}
{\cal L}_{\rm eff} \supset \ & \frac{\lambda'_{ijk}\lambda^{\prime *}_{mnk}}{2m^2_{\tilde{d}_{kR}}} \bigg[\bar\nu_{mL}\gamma^\mu \nu_{iL}\bar{d}_{nL}\gamma_\mu d_{jL}\nonumber \\
& +\bar{e}_{mL}\gamma^\mu e_{iL}\left(\bar u_L V_{\rm CKM}\right)_n \gamma_\mu \left(V_{\rm CKM}^\dag u_L\right)_j \nonumber \\
& -\nu_{mL}\gamma^\mu e_{iL}\bar d_{nL}\gamma_\mu \left(V_{\rm CKM}^\dag u_L\right)_j ~+~\text{h.c.} \bigg] \nonumber \\
& -\frac{\lambda'_{ijk}\lambda^{\prime *}_{mjn}}{2m^2_{\tilde{u}_{jL}}}
\bar e_{mL}\gamma^\mu e_{iL}\bar{d}_{kR}\gamma_\mu d_{nR} ~,
\label{Leff}
\end{align}
where we only show the terms relevant for the following discussion. These operators are of type ${\cal O}_{V_L}$ in~\eqref{eq:vector} and the NP scale $\Lambda$ is given by the squark masses. Note that the NP scale in the first three terms is given by the mass of the sbottom right, while in the last term it is given by the mass of the stop left. These two masses are not necessarily related and we will consider them as independent model parameters.
Naturalness arguments suggest that the stop mass should be below 1~TeV for tuning of O($10^{-2}$); if, on the other hand,  we can tolerate tuning to say
around $\approx 10^{-4}$, the stop mass can be about 10 TeV.
Given the above Lagrangian, the RPV scenario will lead to the collider signals discussed in the previous section with distributions given for the vector case. The operators are SM-like, but the deviation in the distributions in Fig.~\ref{fig:dist} is due to the heavy mediator. 

Since we consider only light third family sfermions and want to explain the $B$-anomalies which requires at least one of the $d$-flavors to be a $b$-quark and the lepton to be of tau flavor in~\eqref{Leff}, we will be interested in the $\lambda^\prime_{3j3}$ (with $j = 1,2,3$  and other indices set to 3).
For simplicity we take all $\lambda^\prime$ couplings as real.
  
\begin{figure*}[t!]
\centering
\includegraphics[width=0.92\textwidth]{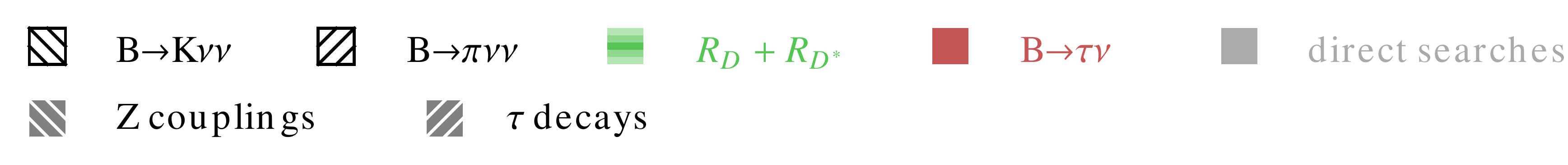} \\[5pt] 
\includegraphics[width=0.46\textwidth]{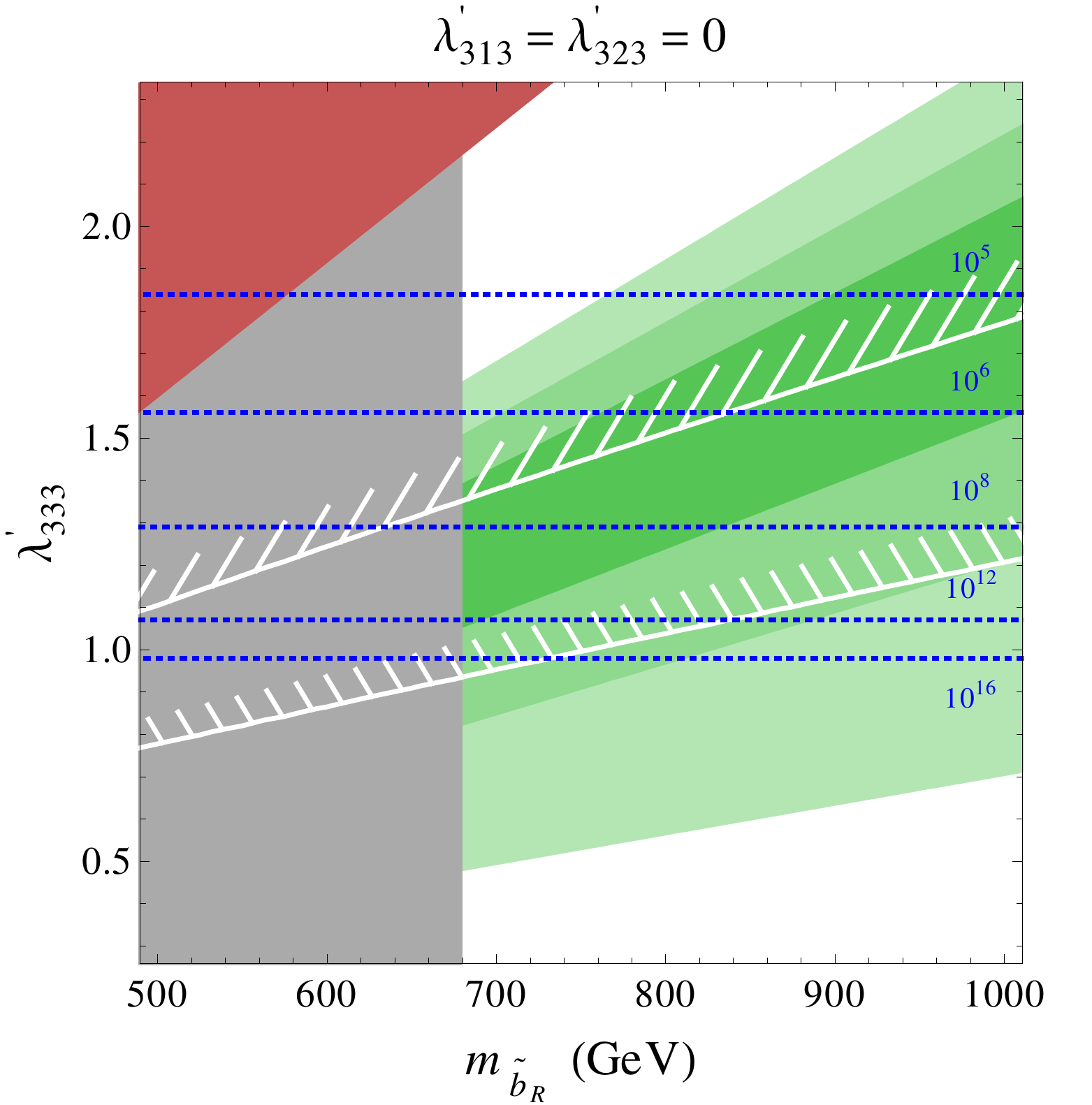} ~~~~~
\includegraphics[width=0.46\textwidth]{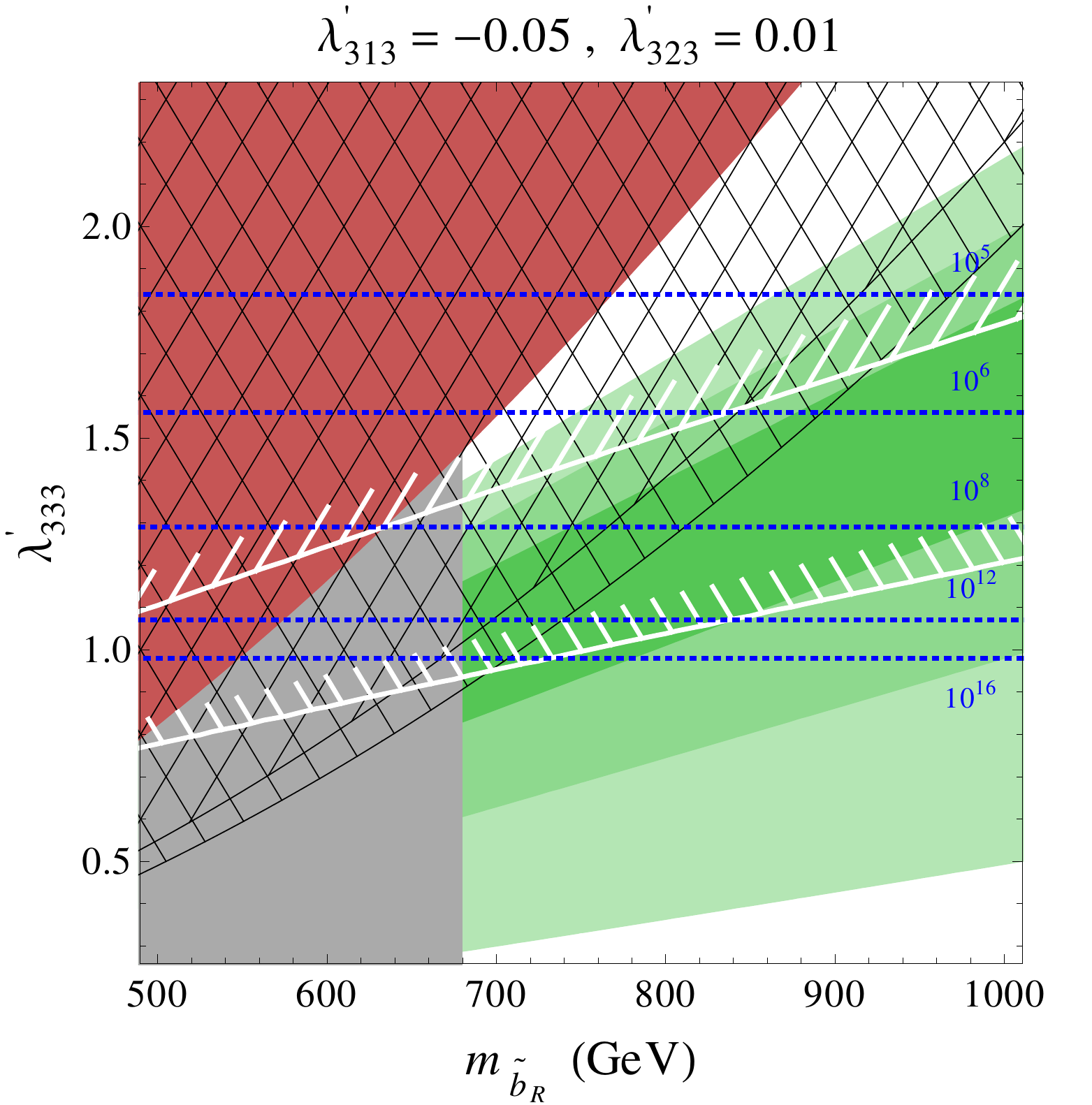} \\[5pt]
\includegraphics[width=0.46\textwidth]{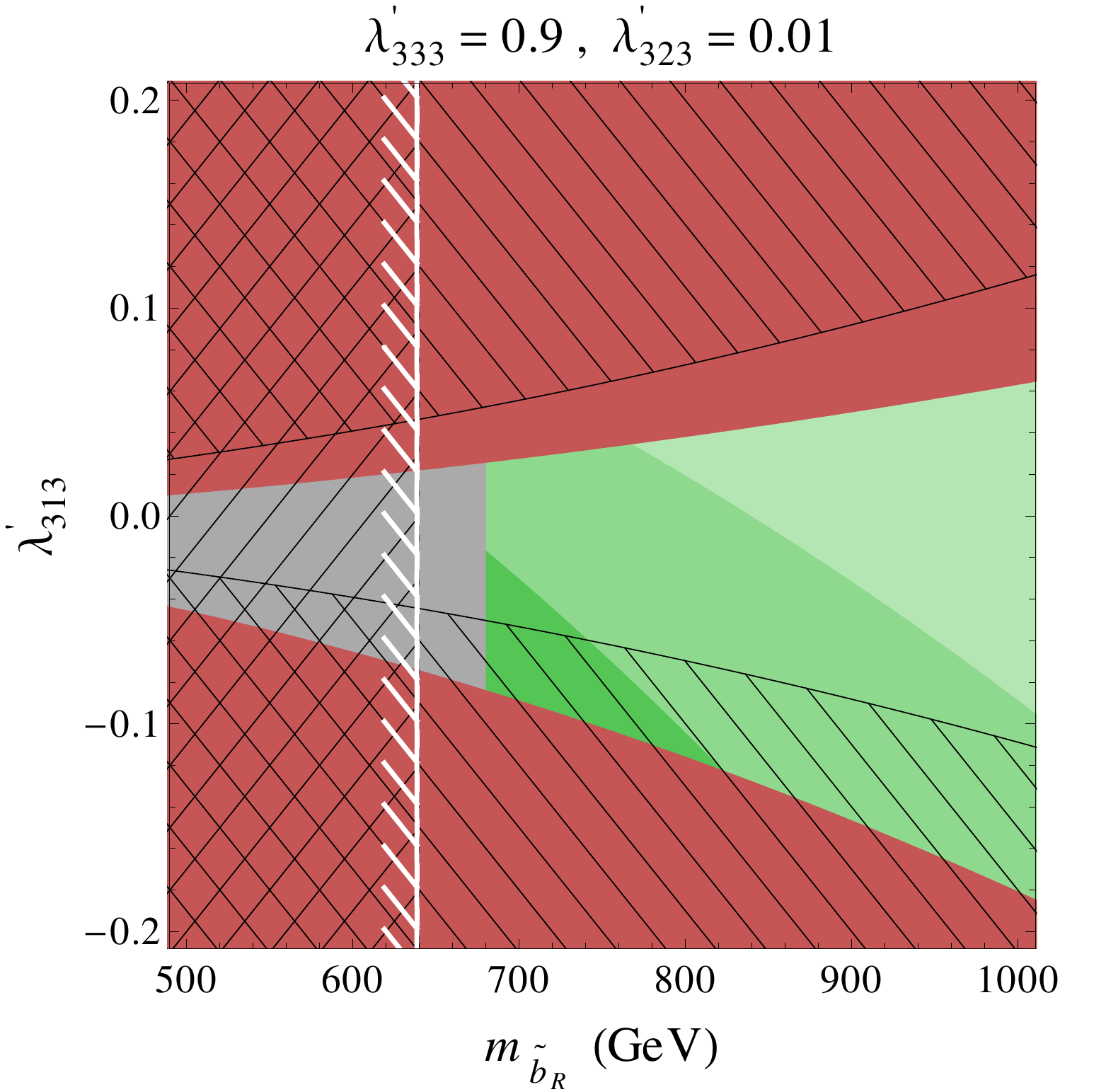} ~~~~~
\includegraphics[width=0.46\textwidth]{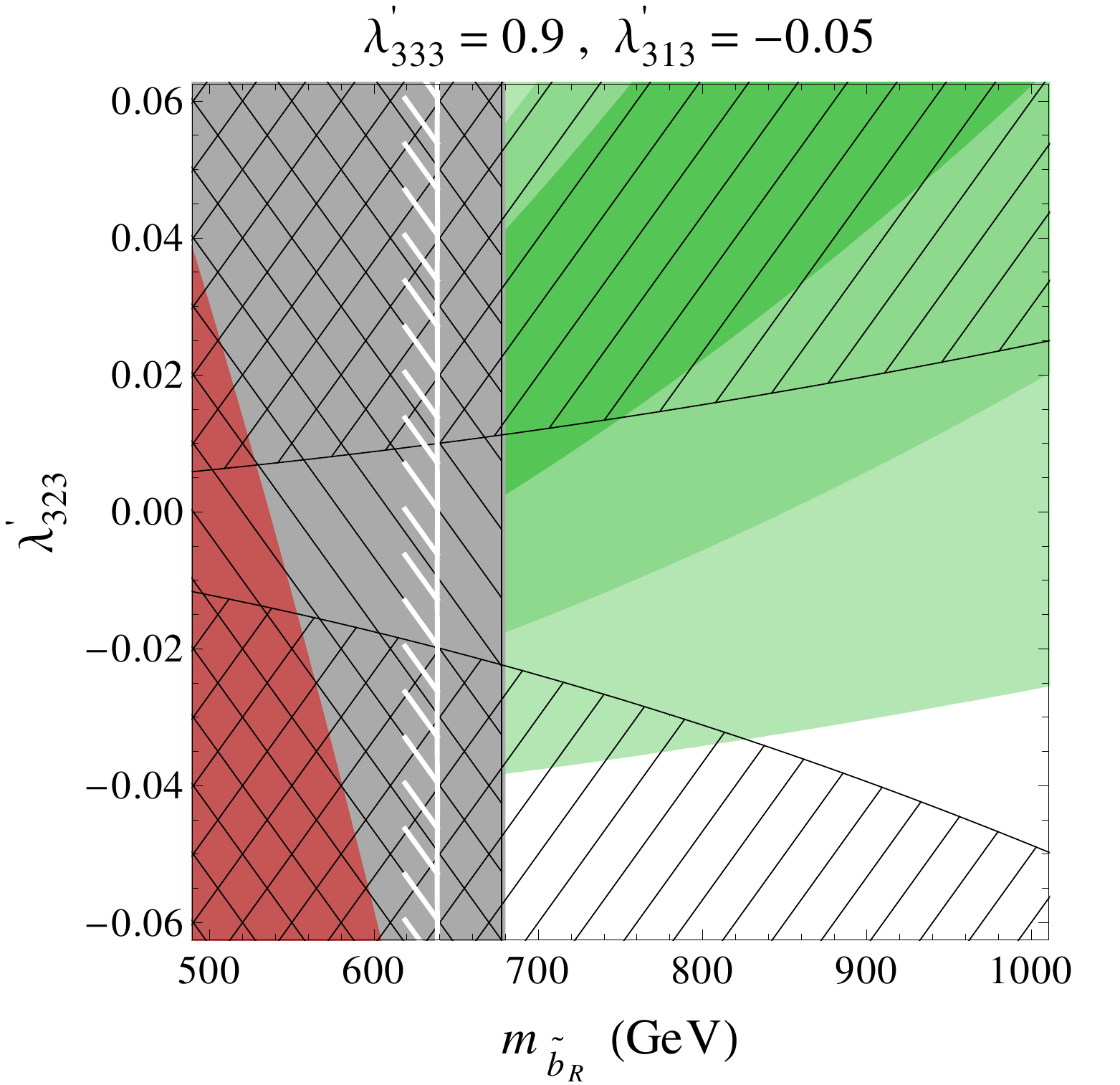} 
\caption{RPV parameter space satisfying the $R_{D^{(*)}}$ anomaly and other relevant constraints.}
\label{fig:RDRDstar}
\end{figure*}

The sbottoms with RPV couplings can contribute to $R_{D^{(*)}}$ at the tree level. 
The corresponding 3rd term in the effective Lagrangian in~(\ref{Leff}) has the same chirality structure as in the SM, implying that the new physics effect is a simple rescaling of the $B \to D^{(*)} \tau\nu$ decay rates. We find
\begin{eqnarray}
 && \frac{R_D}{R_D^\text{SM}} = \frac{R_{D^*}}{R_{D^*}^\text{SM}} = \left| 1 + \frac{v^2}{2 m^2_{\tilde b_R}} X_c  \right|^2 ~, \\
 && X_c= |\lambda_{333}^\prime|^2 + \lambda_{333}^\prime \lambda_{323}^\prime \frac{V_{cs}}{V_{cb}} + \lambda_{333}^\prime \lambda_{313}^\prime \frac{V_{cd}}{V_{cb}} ~,
\end{eqnarray}
where $v = 246$~GeV is the Higgs vev. 
Combining the SM predictions with the experimental world average taking into account the correlated uncertainties, we find the following best fit value in our scenario
\begin{equation} \label{eq:fit}
 \frac{R_D}{R_D^\text{SM}} = \frac{R_{D^*}}{R_{D^*}^\text{SM}} = 1.21 \pm 0.06 ~.
\end{equation}
In Fig.~\ref{fig:RDRDstar} we show in a few benchmark scenarios regions of parameter space in planes of the sbottom mass vs. RPV couplings that can accommodate the discrepant $R_{D^{(*)}}$ results at the 1$\sigma$, 2$\sigma$, and 3$\sigma$ level in green.

Also shown are the most important additional constraints. 
Strongly related to $R_D^{(*)}$ is the decay $B \to \tau \nu$~\cite{Nandi:2016wlp} that also receives tree level contributions from sbottom exchange. We find
\begin{eqnarray} 
 && \frac{\mathcal{B}(B \to \tau \nu)}{\mathcal{B}(B \to \tau\nu)_\text{SM}}  = \left| 1 + \frac{v^2}{2 m^2_{\tilde b_R}} X_u  \right|^2  ~, \\
 && X_u =  |\lambda_{333}^\prime|^2 + \lambda_{333}^\prime \lambda_{323}^\prime \frac{V_{us}}{V_{ub}} + \lambda_{333}^\prime \lambda_{313}^\prime \frac{V_{ud}}{V_{ub}} ~. \label{eq:Btaunu}
\end{eqnarray}
Note that the term proportional to $\lambda_{313}^\prime$ is strongly enhanced by CKM factors. 
The most important parametric input for the SM prediction of $\mathcal{B}(B \to \tau\nu)$ is the $B$ meson decay constant $f_B$ and the absolute value of the CKM matrix element $V_{ub}$.
Using $f_B = (0.191 \pm 0.007)$~GeV from~\cite{Na:2012kp,Aoki:2013ldr} and $V_{ub}^\text{exc.} = (3.61 \pm 0.32) \times 10^{-3}$ from~\cite{Flynn:2015mha}, one finds~\cite{Nandi:2016wlp}
\begin{equation}
 \mathcal{B}(B \to \tau \nu)_\text{SM} =  (0.947 \pm 0.182) \times 10^{-4} ~,
\end{equation}
in good agreement with the experimental average~\cite{Amhis:2016xyh} 
\begin{equation}
 \mathcal{B}(B \to \tau \nu)_\text{exp} = (1.06 \pm 0.19) \times 10^{-4} ~.
\end{equation}
The $2\sigma$ constraint from $\mathcal{B}(B \to \tau\nu)$ is shown in Fig.~\ref{fig:RDRDstar} in red.
As one expects, $B \to \tau\nu$ strongly constrains the coupling $\lambda^\prime_{313}$.
The decay modes $B \to \pi\tau\nu$ and $B \to \rho \tau \nu$ probe the same quark level transition as $B \to \tau\nu$, but we find that they give weaker constraints throughout the interesting parameter space.

Additional important constraints arise from the rare FCNC decays $B \to K \nu\nu$ and $B \to \pi \nu\nu$. In the SM, the branching ratios are strongly suppressed~\cite{Buras:2014fpa,Du:2015tda}
\begin{eqnarray}
 \mathcal{B}(B^+ \to K^+ \nu\nu)_\text{SM} &=& (3.98 \pm 0.43 \pm 0.19) \times 10^{-6} ~, \\
 \mathcal{B}(B^+ \to \pi^+ \nu\nu)_\text{SM} &=& (1.46 \pm 0.14) \times 10^{-7} ~.
\end{eqnarray}
Currently only upper bounds on these branching ratios are available. At 90\% confidence level one has~\cite{Lees:2013kla,Lutz:2013ftz}
\begin{eqnarray}
 \mathcal{B}(B^+ \to K^+ \nu\nu)_\text{exp} &<& 1.6 \times 10^{-5}~, \\
 \mathcal{B}(B^+ \to \pi^+ \nu\nu)_\text{exp} &<& 9.8 \times 10^{-5}~.
\end{eqnarray}
These bounds give strong constraints on the RPV parameters, as the sbottoms with RPV couplings can contribute to the decays already at the tree level. We find
\begin{equation}
 \frac{\mathcal{B}(B \to K \nu\nu)}{\mathcal{B}(B \to K\nu\nu)_\text{SM}} = \frac{2}{3} + \frac{1}{3} \left| 1 + \frac{v^2}{m^2_{\tilde b_R}} \frac{\pi s_W^2}{\alpha_\text{em}} \frac{\lambda_{333}^\prime \lambda_{323}^\prime}{V_{tb} V_{ts}^*} \frac{1}{X_t}  \right|^2  ~,
\end{equation}
\begin{equation}
 \frac{\mathcal{B}(B \to \pi \nu\nu)}{\mathcal{B}(B \to \pi\nu\nu)_\text{SM}} = \frac{2}{3} + \frac{1}{3} \left| 1 + \frac{v^2}{m^2_{\tilde b_R}} \frac{\pi s_W^2}{\alpha_\text{em}} \frac{\lambda_{333}^\prime \lambda_{313}^\prime}{V_{tb} V_{td}^*} \frac{1}{X_t}  \right|^2  ~,
\end{equation}
where $X_t = 1.469\pm0.017$~\cite{Buras:2014fpa} is a SM loop function.
These decays put strong constraints on the $\lambda_{313}^\prime$ and $\lambda_{323}^\prime$ couplings, respectively. This is shown in Fig.~\ref{fig:RDRDstar} by the black hatched regions.

We also consider constraints from direct searches for RPV sbottoms at the LHC. In particular, we consider the pair production of the sbottoms $pp \to \tilde b \tilde b$ followed by decays through the RPV coupling $\lambda_{333}^\prime$, $\tilde b \to t \tau$. We use the the CMS analysis~\cite{Khachatryan:2015bsa} of $pp \to \tau^+\tau^- t\bar t$ to put constraints in the mass-coupling plane. 
We find that this analysis leads to a lower bound on the sbottom mass of $m_{\tilde b_R} \gtrsim 680$~GeV, as shown in the plots of Fig.~\ref{fig:RDRDstar} in gray.

Additional constraints can be obtained from measurements of $Z$ boson couplings and tau decay rates~\cite{Feruglio:2016gvd,Feruglio:2017rjo}.\footnote{We thank Paride Paradisi for pointing this out.} At the loop level, sizable corrections to $Z$ and $W$ couplings involving left-handed tau leptons arise in our setup. The dominant effects come from loops involving sbottoms and top quarks that are proportional to the top quark mass. All other $Z$ and $W$ couplings are not affected significantly. We find
\begin{eqnarray}
 \frac{g_{Z\tau_L\tau_L}}{g_{Z\ell_L\ell_L}} &=& 1 - \frac{3 (\lambda_{333}^\prime)^2}{16\pi^2} \frac{1}{1 - 2 s_W^2} \frac{m_t^2}{m^2_{\tilde b_R}} f_Z\left(\frac{m_t^2}{m^2_{\tilde b_R}}\right) ~, \\ 
 \frac{g_{W\tau_L\nu_\tau}}{g_{W\ell_L\nu_\ell}} &=& 1 - \frac{3 (\lambda_{333}^\prime)^2}{16\pi^2} \frac{1}{4} \frac{m_t^2}{m^2_{\tilde b_R}} f_W\left(\frac{m_t^2}{m^2_{\tilde b_R}}\right) ~,
\end{eqnarray}
where $\ell = e, \mu$, $s_W$ is the sine of the weak mixing angle, and the loop functions are given by $f_Z(x) = \frac{1}{x-1} - \frac{\log(x)}{(x-1)^2}$, $f_W(x) = \frac{1}{x-1} - \frac{(2-x)\log(x)}{(x-1)^2}$.
In the leading log approximation, the above expressions are consistent with the results in~\cite{Feruglio:2016gvd,Feruglio:2017rjo}.
The $Z$ couplings to leptons have been measured at the few permille level at LEP and SLD. Using the results from ~\cite{ALEPH:2005ab}, we profile over the unaffected couplings taking into account the reported error correlations and obtain
\begin{equation} \label{eq:Z}
 \frac{g_{Z\tau_L\tau_L}}{g_{Z\ell_L\ell_L}} = 1.0013 \pm 0.0019 ~.
\end{equation}
The best constraints on the $W$ couplings to taus in our scenario are obtained from measured $tau$ decay rates compared to the muon decay rate. Taking into account the error correlations of measurements of leptonic and semi-hadronic decays reported in~\cite{Amhis:2016xyh}, we find
\begin{equation} \label{eq:W}
 \frac{g_{W\tau_L\nu_\tau}}{g_{W\ell_L\nu_\ell}} = 1.0007 \pm 0.0013 ~.
\end{equation}
The corresponding constraints on the RPV parameter space are shown in the plots of Fig.~\ref{fig:RDRDstar} by the white hatched contours. Parameter space to the top left of the contours is excluded. We find that in particular the $Z$ couplings lead to strong constraints in our scenario.

In addition to the shown constraints, we also considered effects of RPV sbottoms on rare Kaon decays $K \to \pi \nu\nu$, leptonic charm decays $D \to \tau \nu$ and $D_s \to \tau \nu$, and hadronic tau decays $\tau \to K \nu$ and $\tau \to \pi \nu$. 
While these decays probe complementary combinations of the $\lambda^\prime_{3i3}$ couplings, we find that they do not lead to any relevant constraints in the scenarios we are considering.
Also the $B_c$ lifetime~\cite{Li:2016vvp,Alonso:2016oyd} does not lead to relevant bounds as the contributions from the RPV sbottoms to $B_c \to \tau \nu$ are not chirally enhanced with respect to the SM. We also
checked that the remaining parameter space shown in Fig.~\ref{fig:RDRDstar} is consistent with other constraints on RPV couplings from various low-energy precision observables~\cite{Kao:2009fg}. 

In our setup, we generically also expect NP effects in processes like $b \bar b \to \tau \tau$ and $b \to s \tau \tau$, coming from the tree level exchange of stops. The corresponding couplings of the stop to bottom and leptons are related by SUSY to those of the sbottom with tops and leptons. The stop and sbottom masses, however, are independent parameters. The results of~\cite{Faroughy:2016osc} indicate that for a coupling of $\lambda'_{333} \sim 2$, the $b \bar b \to \tau \tau$ process constrains the stop mass around $m_{\tilde t} \gtrsim 500$~GeV. Constraints from $b \to s \tau \tau$ are considerably weaker than the corresponding ones from $b \to s \nu \nu$ decays, given the poor experimental sensitivity to the decays with taus in the final state.

Overall, to explain the $R_{D^{(*)}}$ anomaly at $1\sigma$, large values of $\lambda_{333}^\prime \sim 1 - 2$ are required for sbottom masses compatible with direct LHC searches. 
For such large $\lambda_{333}^\prime$ at the TeV scale, this coupling develops a Landau pole below the GUT scale. In Fig.~\ref{fig:RDRDstar}, the position of the Landau pole in GeV is indicated by the dotted blue lines, which are obtained by numerically solving the coupled system of 1-loop RGEs of the $\lambda_{333}^\prime$ coupling~\cite{Allanach:1999mh}, the top Yukawa, and the three gauge couplings in the presence of one generation of light sfermions.
Perturbativity up to the GUT scale requires $\lambda_{333}^\prime \lesssim 1$.
Also the $Z$ coupling constraints limit effects in $R_{D^{(*)}}$. In the viable parameter space the $R_{D^{(*)}}$ anomaly can reasonably well be resolved especially in view of the largish experimental errors that exist
up to now.

\begin{figure}[t!]
\centering
\includegraphics[width=0.46\textwidth]{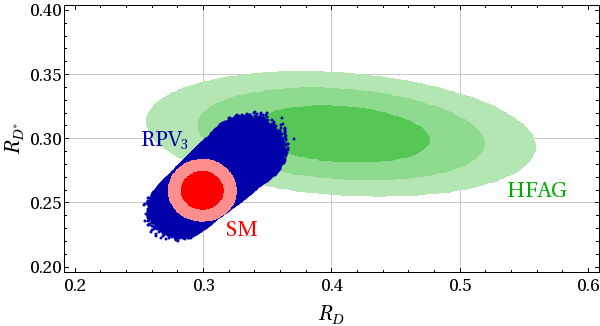} 
\caption{The SM predictions (red), experimental world average (green), and accessible values in our RPV-SUSY scenario (blue) in the $R_D$ vs. $R_{D^*}$ plane. For the SM, bearing in mind recent works~\cite{ Lattice:2015rga,Bernlochner:2017jka, Bigi:2017jbd} we are taking ($R_D^\text{SM},R_{D^*}^\text{SM})=(0.299\pm0.011,0.260\pm0.010$).
}
\label{fig:RDvsRDstar}
\end{figure}

In Fig.~\ref{fig:RDvsRDstar} we map the allowed regions of parameter space into the $R_D$ vs. $R_{D^*}$ plane.
This mapping is possible as the NP effect is a simple rescaling of the SM and efficiencies remain SM-like. 
The red region shows the SM predictions at $1$ and $2\sigma$, $R_D^\text{SM} = 0.299 \pm 0.011$ [cf.~\eqref{RDRDstarSM}] and $R_{D^*}^\text{SM} = 0.260 \pm 0.010$ with zero error correlation. For $R_{D^*}^\text{SM}$ we take the central value to be the average of~\cite{Bernlochner:2017jka}
and~\cite{Bigi:2017jbd} but the error to be the full spread
between~\cite{Bigi:2017jbd} and a previous determination~\cite{Fajfer:2012vx}.
In green we show the experimental world average from~\cite{Amhis:2016xyh} at $1, 2$, and $3\sigma$.
The blue region spans $R_D=(0.254,0.371)$ and $R_{D^*}=(0.220,0.320)$ and shows values that can be obtained in our setup consistent with all above-mentioned constraints.
To obtain this region we scan the sbottom mass between the lower experimental bound of $m_{\tilde b} > 680$~GeV up to $m_{\tilde b} < 1$~TeV. The RPV couplings are varied in the ranges $0 < \lambda'_{333} < 2$, $-0.1 < \lambda'_{323} < 0.1$, and $-0.3 < \lambda'_{313} < 0.3$. We impose all the constraints discussed above. The blue points correspond to RPV couplings that remain perturbative up to the GUT scale. Relaxing this requirement and allowing the $\lambda_{333}^\prime$ coupling to develop a Landau pole before the GUT scale does not lead to larger effects in $R_{D^{(*)}}$, given the constraints from $Z$ couplings.

\section{Discussion}

The same RPV couplings that generate the desired effects in $R_{D^{(*)}}$ also generate a neutrino mass through the bottom-sbottom loop~\cite{Barbier:2004ez}:
\begin{align}\label{eq:neutrinomass} 
\Delta M_{\nu, ij}^{\lambda'} \ \simeq \ \frac{3}{8\pi^2}\frac{m_b^2(A_b-\mu \tan\beta)}{m_{\tilde{b}}^2}\lambda'_{i33}\lambda'_{j33} ~.
\end{align}
For $m_{\tilde{b}}\sim 1$ TeV and $\lambda'_{333}\sim {\cal O}(1)$, we get $m_\nu\sim 0.1$ MeV. This contribution could be avoided if the trilinear coupling $A_b$ and the term $\mu \tan\beta$ in (\ref{eq:neutrinomass}) cancel each other precisely. Another option to get sub-eV scale neutrino masses is to invoke cancellations between the $\lambda'$ induced contributions and other unrelated contributions to neutrino masses. These additional contributions could either arise at the tree level or the loop level. Tree level contributions can originate e.g. from a standard see-saw mechanism with heavy right-handed neutrinos, or from neutralino-neutrino mixing due to bilinear RPV terms. At the loop level additional $\lambda_{ijk} L_iL_jE_k^c$ terms in the RPV Lagrangian could contribute to neutrino masses, e.g.
\begin{align}
M_{\nu, ij}^{\lambda} \ \simeq \ \frac{1}{8\pi^2}\frac{m_\tau^2(A_\tau-\mu \tan\beta)}{m_{\tilde{\tau}}^2}\lambda_{i33}\lambda_{j33} ~.
\end{align}
The stau mass and the $\lambda_{k33}$ couplings can be chosen such that $\Delta M_{\nu, ij}^{\lambda} = -\Delta M_{\nu, ij}^{\lambda'}$.
Note that appropriately chosen $\lambda_{ijk}$ couplings could also explain the $(g-2)_\mu$ anomaly~\cite{Chakraborty:2015bsk}. Also note that the RPV couplings $\lambda''_{ijk}U_i^cD_j^cD_k^c$ should be explicitly forbidden, e.g. by imposing baryon triality~\cite{Ibanez:1991hv}, to avoid rapid proton decay~\cite{Smirnov:1996bg, Nath:2006ut}.

We would also like to comment on the possibilities to address another hint for LFUV in the rare $B$ meson decays based on the $b \to s \ell^+\ell^-$ transition. The LHCb collaboration measured the ratios
\begin{equation}
 R_K = \frac{\mathcal B(B \to K \mu^+\mu^-)}{\mathcal B(B \to K e^+e^-)}
\,,~
R_{K^*} = \frac{\mathcal B(B \to K^* \mu^+\mu^-)}{\mathcal B(B \to K^* e^+e^-)} \,.
\end{equation}
and finds~\cite{Aaij:2014ora,Aaij:2017vbb}
\begin{align}
 R_K^{[1,6]} &= 0.745 ^{+0.090}_{-0.074} \pm 0.036 ~, \\
 R_{K^*}^{[0.045, 1.1]} &= 0.66^{+0.11}_{-0.07} \pm 0.03\,,\\
 R_{K^*}^{[1.1, 6]} &= 0.69^{+0.11}_{-0.07} \pm 0.05\,,
\end{align}
where the superscript corresponds to the di-lepton invariant mass bin in GeV$^2$.
The SM predictions for $R_K^{[1,6]}$ and $R_{K^*}^{[1.1,6]}$ are 1 with percent level accuracy~\cite{Bordone:2016gaq}. 
The SM prediction for the low di-lepton invariant mass bin $R_{K^*}^{[0.045,1.1]}$ is slightly below 1, mainly due to phase space effects.

Our RPV setup allows for two qualitatively different contributions to $b \to s \ell^+\ell^-$ decays:
at tree level one finds contributions from integrating out the left-handed stop;
at the 1-loop level also the right-handed sbottom can contribute, in analogy to the scalar leptoquark considered in~\cite{Bauer:2015knc}.
The tree level contribution is captured in the effective Lagrangian~(\ref{Leff}). It contains a right-handed down-type quark current and therefore predicts that a suppressed $R_K$ is correlated with an enhanced 
$R_{K^*}$ and vice versa~\cite{Hiller:2014ula}, in conflict with 
the findings by LHCb. 
The 1-loop contribution has the correct chirality structure and can in principle suppress both $R_K$ and $R_{K^*}$. However, it has been pointed out in~\cite{Deshpand:2016cpw,Becirevic:2016oho} that the loop contribution typically does not give appreciable effects in $b \to s \ell \ell$ transitions once constraints from other low energy data are taken into account.

A detailed study would be required to ascertain whether or not 
the discussed RPV SUSY framework contains additional contributions to 
$b \to s \ell \ell$ that could simultaneously accommodate all the hints 
for LFUV in rare $B$ decays.

\section{Conclusion}

$B$-physics experiments have reported appreciable ($\approx~4\sigma$) 
deviations in the tree-level observables, $R_{D^{(*)}}$.  We have proposed a model-independent collider signal of the form $pp \to b \tau \bar{\nu}_{\tau}$, $pp \to b \ell \bar{\nu}_{\ell}$ that should be searched for at the LHC as soon as possible to verify the anomaly.
Taking the reported deviations from the SM at face value, we also point out that there is the exciting possibility that the origin of the anomalies might be related to the naturalness of the Higgs boson in the SM. In particular, we discussed a minimal (i.e.~involving only the third generation), effective RPV-SUSY scenario as the underlying cause, and identified the parameter space of the RPV couplings and sbottom mass that could explain the $R_{D^{(*)}}$ anomaly, while being consistent with other experimental constraints, as well as preserving the gauge coupling unification.

\section{Acknowledgments} 

We thank Claude Bernard, Paride Paradisi, Dean Robinson, Brian Shuve, and Florian Staub for useful comments and discussions. W.A. acknowledges financial support by the University of Cincinnati. The work of A.S. is supported in part by the US DOE Contract No. DE-SC 0012704.



\begin{thebibliography}{99}

\bibitem{Hirose:2017vbz} 
  S.~Hirose [Belle Collaboration],
  arXiv:1705.05100 [hep-ex].


\bibitem{talk2}
G.~Wormser [LHCb Collaboration],
Talk presented at Moriond EW 2017.

\bibitem{Kiers:1997zt} 
  K.~Kiers and A.~Soni,
  Phys.\ Rev.\ D {\bf 56}, 5786 (1997)
  [hep-ph/9706337].

\bibitem{Chen:2006nua} 
  C.~H.~Chen and C.~Q.~Geng,
  JHEP {\bf 0610}, 053 (2006)
  [hep-ph/0608166].

\bibitem{Nierste:2008qe} 
  U.~Nierste, S.~Trine and S.~Westhoff,
  Phys.\ Rev.\ D {\bf 78}, 015006 (2008)
  [arXiv:0801.4938 [hep-ph]].

\bibitem{Kamenik:2008tj} 
  J.~F.~Kamenik and F.~Mescia,
  Phys.\ Rev.\ D {\bf 78}, 014003 (2008)
  [arXiv:0802.3790 [hep-ph]].

\bibitem{Lees:2012xj} 
  J.~P.~Lees {\it et al.} [BaBar Collaboration],
  Phys.\ Rev.\ Lett.\  {\bf 109}, 101802 (2012)
  [arXiv:1205.5442 [hep-ex]].
  
\bibitem{Lees:2013uzd} 
  J.~P.~Lees {\it et al.} [BaBar Collaboration],
  Phys.\ Rev.\ D {\bf 88}, no. 7, 072012 (2013)
  [arXiv:1303.0571 [hep-ex]].

\bibitem{Huschle:2015rga} 
  M.~Huschle {\it et al.} [Belle Collaboration],
  Phys.\ Rev.\ D {\bf 92}, no. 7, 072014 (2015)
  [arXiv:1507.03233 [hep-ex]].

\bibitem{Sato:2016svk} 
  Y.~Sato {\it et al.} [Belle Collaboration],
  Phys.\ Rev.\ D {\bf 94}, no. 7, 072007 (2016)
  [arXiv:1607.07923 [hep-ex]].

\bibitem{Abdesselam:2016xqt} 
  A.~Abdesselam {\it et al.} [Belle Collaboration],
  arXiv:1608.06391 [hep-ex].

\bibitem{Hirose:2016wfn} 
  S.~Hirose {\it et al.} [Belle Collaboration],
  arXiv:1612.00529 [hep-ex].

\bibitem{Hirose:2017dxl} 
  S.~Hirose {\it et al.} [Belle Collaboration],
  arXiv:1709.00129 [hep-ex].
  
\bibitem{Aaij:2015yra} 
  R.~Aaij {\it et al.} [LHCb Collaboration],
  Phys.\ Rev.\ Lett.\  {\bf 115}, no. 11, 111803 (2015)
  Addendum: [Phys.\ Rev.\ Lett.\  {\bf 115}, no. 15, 159901 (2015)]
  [arXiv:1506.08614 [hep-ex]].
  
\bibitem{new} G. Wormser [LHCb Collaboration], Talk presented at FPCP 2017 Conference, Prague.  

\bibitem{Amhis:2016xyh} 
  Y.~Amhis {\it et al.},
  arXiv:1612.07233 [hep-ex] and online update at \verb|http://www.slac.stanford.edu/xorg/hfag|
  
\bibitem{Bernlochner:2017jka} 
  F.~U.~Bernlochner, Z.~Ligeti, M.~Papucci and D.~J.~Robinson,
  arXiv:1703.05330 [hep-ph].

\bibitem{Fajfer:2012vx} 
  S.~Fajfer, J.~F.~Kamenik and I.~Nisandzic,
  Phys.\ Rev.\ D {\bf 85}, 094025 (2012)
  [arXiv:1203.2654 [hep-ph]].

\bibitem{Bigi:2016mdz} 
  D.~Bigi and P.~Gambino,
  Phys.\ Rev.\ D {\bf 94}, no. 9, 094008 (2016)
  [arXiv:1606.08030 [hep-ph]].

\bibitem{Lattice:2015rga} 
  J.~A.~Bailey {\it et al.} [MILC Collaboration],
  Phys.\ Rev.\ D {\bf 92}, no. 3, 034506 (2015)
  [arXiv:1503.07237 [hep-lat]].

\bibitem{Na:2015kha} 
  H.~Na {\it et al.} [HPQCD Collaboration],
  Phys.\ Rev.\ D {\bf 92}, no. 5, 054510 (2015)
  Erratum: [Phys.\ Rev.\ D {\bf 93}, no. 11, 119906 (2016)]
  [arXiv:1505.03925 [hep-lat]].

\bibitem{Bigi:2017jbd} 
  D.~Bigi, P.~Gambino and S.~Schacht,
  arXiv:1707.09509 [hep-ph].
  
\bibitem{Aoki:2016frl} 
  S.~Aoki {\it et al.},
  arXiv:1607.00299 [hep-lat].

\bibitem{Bailey:2014tva} 
  J.~A.~Bailey {\it et al.} [Fermilab Lattice and MILC Collaborations],
  Phys.\ Rev.\ D {\bf 89}, no. 11, 114504 (2014)
  [arXiv:1403.0635 [hep-lat]].
  
\bibitem{Fajfer:2012jt} 
  S.~Fajfer, J.~F.~Kamenik, I.~Nisandzic and J.~Zupan,
  Phys.\ Rev.\ Lett.\  {\bf 109}, 161801 (2012)
  [arXiv:1206.1872 [hep-ph]].

\bibitem{Datta:2012qk} 
  A.~Datta, M.~Duraisamy and D.~Ghosh,
  Phys.\ Rev.\ D {\bf 86}, 034027 (2012)
  [arXiv:1206.3760 [hep-ph]].

\bibitem{Bailey:2012jg} 
  J.~A.~Bailey {\it et al.},
  Phys.\ Rev.\ Lett.\  {\bf 109}, 071802 (2012)
  [arXiv:1206.4992 [hep-ph]].

\bibitem{Tanaka:2012nw} 
  M.~Tanaka and R.~Watanabe,
  Phys.\ Rev.\ D {\bf 87}, no. 3, 034028 (2013)
  [arXiv:1212.1878 [hep-ph]].
  
\bibitem{Biancofiore:2013ki} 
  P.~Biancofiore, P.~Colangelo and F.~De Fazio,
  Phys.\ Rev.\ D {\bf 87}, no. 7, 074010 (2013)
  [arXiv:1302.1042 [hep-ph]].

\bibitem{Alonso:2015sja} 
  R.~Alonso, B.~Grinstein and J.~Martin Camalich,
  JHEP {\bf 1510}, 184 (2015)
  [arXiv:1505.05164 [hep-ph]].
  
\bibitem{Greljo:2015mma} 
  A.~Greljo, G.~Isidori and D.~Marzocca,
  JHEP {\bf 1507}, 142 (2015)
  [arXiv:1506.01705 [hep-ph]].
  
\bibitem{Calibbi:2015kma} 
  L.~Calibbi, A.~Crivellin and T.~Ota,
  Phys.\ Rev.\ Lett.\  {\bf 115}, 181801 (2015)
  [arXiv:1506.02661 [hep-ph]].
  
\bibitem{Freytsis:2015qca} 
  M.~Freytsis, Z.~Ligeti and J.~T.~Ruderman,
  Phys.\ Rev.\ D {\bf 92}, no. 5, 054018 (2015)
  [arXiv:1506.08896 [hep-ph]].

\bibitem{Bhattacharya:2015ida} 
  S.~Bhattacharya, S.~Nandi and S.~K.~Patra,
  Phys.\ Rev.\ D {\bf 93}, no. 3, 034011 (2016)
  [arXiv:1509.07259 [hep-ph]].

\bibitem{Alonso:2016gym} 
  R.~Alonso, A.~Kobach and J.~Martin Camalich,
  Phys.\ Rev.\ D {\bf 94}, no. 9, 094021 (2016)
  [arXiv:1602.07671 [hep-ph]].
  
\bibitem{Nandi:2016wlp} 
  S.~Nandi, S.~K.~Patra and A.~Soni,
  arXiv:1605.07191 [hep-ph].

\bibitem{Ivanov:2016qtw} 
  M.~A.~Ivanov, J.~G.~Korner and C.~T.~Tran,
  Phys.\ Rev.\ D {\bf 94}, no. 9, 094028 (2016)
  [arXiv:1607.02932 [hep-ph]].
  
\bibitem{Ligeti:2016npd} 
  Z.~Ligeti, M.~Papucci and D.~J.~Robinson,
  JHEP {\bf 1701}, 083 (2017)
  [arXiv:1610.02045 [hep-ph]].
  
\bibitem{Bardhan:2016uhr} 
  D.~Bardhan, P.~Byakti and D.~Ghosh,
  JHEP {\bf 1701}, 125 (2017)
  [arXiv:1610.03038 [hep-ph]].
  
\bibitem{Bhattacharya:2016zcw} 
  S.~Bhattacharya, S.~Nandi and S.~K.~Patra,
  arXiv:1611.04605 [hep-ph].
  
\bibitem{Alonso:2016oyd} 
  R.~Alonso, B.~Grinstein and J.~Martin Camalich,
  Phys.\ Rev.\ Lett.\  {\bf 118}, no. 8, 081802 (2017)
  [arXiv:1611.06676 [hep-ph]].

\bibitem{Celis:2016azn} 
  A.~Celis, M.~Jung, X.~Q.~Li and A.~Pich,
  arXiv:1612.07757 [hep-ph].

\bibitem{Bordone:2017anc} 
  M.~Bordone, G.~Isidori and S.~Trifinopoulos,
  arXiv:1702.07238 [hep-ph].
  
\bibitem{Bauer:2015knc} 
  M.~Bauer and M.~Neubert,
  Phys.\ Rev.\ Lett.\  {\bf 116}, no. 14, 141802 (2016)
  [arXiv:1511.01900 [hep-ph]].

\bibitem{Hati:2015awg} 
  C.~Hati, G.~Kumar and N.~Mahajan,
  JHEP {\bf 1601}, 117 (2016)
  [arXiv:1511.03290 [hep-ph]].

\bibitem{Fajfer:2015ycq} 
  S.~Fajfer and N.~Kosnik,
  Phys.\ Lett.\ B {\bf 755}, 270 (2016)
  [arXiv:1511.06024 [hep-ph]].
  
\bibitem{Barbieri:2015yvd} 
  R.~Barbieri, G.~Isidori, A.~Pattori and F.~Senia,
  Eur.\ Phys.\ J.\ C {\bf 76}, no. 2, 67 (2016)
  [arXiv:1512.01560 [hep-ph]].
  
\bibitem{Cline:2015lqp} 
  J.~M.~Cline,
  Phys.\ Rev.\ D {\bf 93}, no. 7, 075017 (2016)
  [arXiv:1512.02210 [hep-ph]].
  
\bibitem{Zhu:2016xdg} 
  J.~Zhu, H.~M.~Gan, R.~M.~Wang, Y.~Y.~Fan, Q.~Chang and Y.~G.~Xu,
  Phys.\ Rev.\ D {\bf 93}, no. 9, 094023 (2016)
  [arXiv:1602.06491 [hep-ph]].
  
\bibitem{Boucenna:2016wpr} 
  S.~M.~Boucenna, A.~Celis, J.~Fuentes-Martin, A.~Vicente and J.~Virto,
  Phys.\ Lett.\ B {\bf 760}, 214 (2016)
  [arXiv:1604.03088 [hep-ph]].
  
\bibitem{Das:2016vkr} 
  D.~Das, C.~Hati, G.~Kumar and N.~Mahajan,
  Phys.\ Rev.\ D {\bf 94}, 055034 (2016)
  [arXiv:1605.06313 [hep-ph]].
  
\bibitem{Li:2016vvp} 
  X.~Q.~Li, Y.~D.~Yang and X.~Zhang,
  JHEP {\bf 1608}, 054 (2016)
  [arXiv:1605.09308 [hep-ph]].

\bibitem{Boucenna:2016qad} 
  S.~M.~Boucenna, A.~Celis, J.~Fuentes-Martin, A.~Vicente and J.~Virto,
  JHEP {\bf 1612}, 059 (2016)
  [arXiv:1608.01349 [hep-ph]].
  
\bibitem{Deshpand:2016cpw} 
  N.~G.~Deshpande and X.~G.~He,
   Eur.\ Phys.\ J.\ C {\bf 77}, no. 2, 134 (2017)
  [arXiv:1608.04817 [hep-ph]].
  
\bibitem{Becirevic:2016yqi} 
  D.~Becirevic, S.~Fajfer, N.~Kosnik and O.~Sumensari,
  Phys.\ Rev.\ D {\bf 94}, no. 11, 115021 (2016)
  [arXiv:1608.08501 [hep-ph]].
  
\bibitem{Sahoo:2016pet} 
  S.~Sahoo, R.~Mohanta and A.~K.~Giri,
  Phys.\ Rev.\ D {\bf 95}, no. 3, 035027 (2017)
  [arXiv:1609.04367 [hep-ph]].
  
\bibitem{Hiller:2016kry} 
  G.~Hiller, D.~Loose and K.~Schonwald,
  JHEP {\bf 1612}, 027 (2016)
  [arXiv:1609.08895 [hep-ph]].
  
\bibitem{Bhattacharya:2016mcc} 
  B.~Bhattacharya, A.~Datta, J.~P.~Guevin, D.~London and R.~Watanabe,
  JHEP {\bf 1701}, 015 (2017)
  [arXiv:1609.09078 [hep-ph]].
  
\bibitem{Wang:2016ggf} 
  L.~Wang, J.~M.~Yang and Y.~Zhang,
  arXiv:1610.05681 [hep-ph].
  
\bibitem{Popov:2016fzr} 
  O.~Popov and G.~A.~White,
  arXiv:1611.04566 [hep-ph].
  
\bibitem{Barbieri:2016las} 
  R.~Barbieri, C.~W.~Murphy and F.~Senia,
  Eur.\ Phys.\ J.\ C {\bf 77}, no. 1, 8 (2017)
  [arXiv:1611.04930 [hep-ph]].
  
\bibitem{Wei:2017ago} 
  M.~Wei and Y.~Chong-Xing,
  Phys.\ Rev.\ D {\bf 95}, no. 3, 035040 (2017)
  [arXiv:1702.01255 [hep-ph]].
  
\bibitem{Cvetic:2017gkt} 
  G.~Cvetic, F.~Halzen, C.~S.~Kim and S.~Oh,
  arXiv:1702.04335 [hep-ph].

\bibitem{Ko:2017lzd} 
  P.~Ko, Y.~Omura, Y.~Shigekami and C.~Yu,
  arXiv:1702.08666 [hep-ph].
  
\bibitem{Chen:2017hir} 
  C.~H.~Chen, T.~Nomura and H.~Okada,
  arXiv:1703.03251 [hep-ph].

\bibitem{Chen:2017eby} 
  C.~H.~Chen and T.~Nomura,
  arXiv:1703.03646 [hep-ph].
  
\bibitem{Megias:2017ove} 
  E.~Megias, M.~Quiros and L.~Salas,
  arXiv:1703.06019 [hep-ph].
  
\bibitem{Crivellin:2017zlb} 
  A.~Crivellin, D.~Muller and T.~Ota,
  arXiv:1703.09226 [hep-ph].
 
\bibitem{Autermann:2016les} 
  C.~Autermann,
  Prog.\ Part.\ Nucl.\ Phys.\  {\bf 90}, 125 (2016)
  [arXiv:1609.01686 [hep-ex]].
  
\bibitem{Faroughy:2016osc} 
  D.~A.~Faroughy, A.~Greljo and J.~F.~Kamenik,
  Phys.\ Lett.\ B {\bf 764}, 126 (2017)
  [arXiv:1609.07138 [hep-ph]].
  
\bibitem{Chatrchyan:2012jua} 
  S.~Chatrchyan {\it et al.} [CMS Collaboration],
  JINST {\bf 8}, P04013 (2013)
  [arXiv:1211.4462 [hep-ex]].

\bibitem{Aad:2009wy} 
  G.~Aad {\it et al.} [ATLAS Collaboration],
  arXiv:0901.0512 [hep-ex].

\bibitem{ECFA} See e.g. talks at 3rd ECFA High Luminosity LHC Experiments Workshop, Aix-Les-Bains, France (2016) [\url{https://indico.cern.ch/event/524795/timetable/}].
  
\bibitem{Alwall:2014hca} 
  J.~Alwall {\it et al.},
  JHEP {\bf 1407}, 079 (2014)
  [arXiv:1405.0301 [hep-ph]].
  
\bibitem{Aad:2012cia} 
  G.~Aad {\it et al.} [ATLAS Collaboration],
  Eur.\ Phys.\ J.\ C {\bf 72}, 2062 (2012)
  [arXiv:1204.6720 [hep-ex]].
  
\bibitem{Ivanov:2017mrj} 
  M.~A.~Ivanov, J.~G.~Korner and C.~T.~Tran,
  Phys.\ Rev.\ D {\bf 95}, no. 3, 036021 (2017)
  [arXiv:1701.02937 [hep-ph]].
 
\bibitem{Alonso:2017ktd} 
  R.~Alonso, J.~Martin Camalich and S.~Westhoff,
  arXiv:1702.02773 [hep-ph].
  
\bibitem{Brust:2011tb} 
  C.~Brust, A.~Katz, S.~Lawrence and R.~Sundrum,
  JHEP {\bf 1203}, 103 (2012)
  [arXiv:1110.6670 [hep-ph]].
  
\bibitem{Papucci:2011wy} 
  M.~Papucci, J.~T.~Ruderman and A.~Weiler,
  JHEP {\bf 1209}, 035 (2012)
  [arXiv:1110.6926 [hep-ph]].
  
\bibitem{Allanach:1999mh} 
  B.~C.~Allanach, A.~Dedes and H.~K.~Dreiner,
  Phys.\ Rev.\ D {\bf 60}, 056002 (1999)
  Erratum: [Phys.\ Rev.\ D {\bf 86}, 039906 (2012)]
  [hep-ph/9902251].
  
\bibitem{Box:2007ss} 
  A.~D.~Box and X.~Tata,
  Phys.\ Rev.\ D {\bf 77}, 055007 (2008)
  Erratum: [Phys.\ Rev.\ D {\bf 82}, 119904 (2010)]
  [arXiv:0712.2858 [hep-ph]].
  
\bibitem{Barbieri:1987fn} 
  R.~Barbieri and G.~F.~Giudice,
  Nucl.\ Phys.\ B {\bf 306}, 63 (1988).

\bibitem{Buckley:2016tbs} 
  M.~R.~Buckley, A.~Monteux and D.~Shih,
  arXiv:1611.05873 [hep-ph].

\bibitem{Deshpande:2012rr} 
  N.~G.~Deshpande and A.~Menon,
  JHEP {\bf 1301}, 025 (2013)
  [arXiv:1208.4134 [hep-ph]].

\bibitem{Na:2012kp} 
  H.~Na, C.~J.~Monahan, C.~T.~H.~Davies, R.~Horgan, G.~P.~Lepage and J.~Shigemitsu,
  Phys.\ Rev.\ D {\bf 86}, 034506 (2012)
  [arXiv:1202.4914 [hep-lat]].
  
\bibitem{Aoki:2013ldr} 
  S.~Aoki {\it et al.},
  Eur.\ Phys.\ J.\ C {\bf 74}, 2890 (2014)
  [arXiv:1310.8555 [hep-lat]].
  
\bibitem{Flynn:2015mha} 
  J.~M.~Flynn, T.~Izubuchi, T.~Kawanai, C.~Lehner, A.~Soni, R.~S.~Van de Water and O.~Witzel,
  Phys.\ Rev.\ D {\bf 91}, no. 7, 074510 (2015)
  [arXiv:1501.05373 [hep-lat]].
  
\bibitem{Buras:2014fpa} 
  A.~J.~Buras, J.~Girrbach-Noe, C.~Niehoff and D.~M.~Straub,
  JHEP {\bf 1502}, 184 (2015)
  [arXiv:1409.4557 [hep-ph]].

\bibitem{Du:2015tda} 
  D.~Du, A.~X.~El-Khadra, S.~Gottlieb, A.~S.~Kronfeld, J.~Laiho, E.~Lunghi, R.~S.~Van de Water and R.~Zhou,
  Phys.\ Rev.\ D {\bf 93}, no. 3, 034005 (2016)
  [arXiv:1510.02349 [hep-ph]].

\bibitem{Lutz:2013ftz} 
  O.~Lutz {\it et al.} [Belle Collaboration],
  Phys.\ Rev.\ D {\bf 87}, no. 11, 111103 (2013)
  [arXiv:1303.3719 [hep-ex]].
  
\bibitem{Lees:2013kla} 
  J.~P.~Lees {\it et al.} [BaBar Collaboration],
  Phys.\ Rev.\ D {\bf 87}, no. 11, 112005 (2013)
  [arXiv:1303.7465 [hep-ex]].

\bibitem{Khachatryan:2015bsa} 
  V.~Khachatryan {\it et al.} [CMS Collaboration],
  JHEP {\bf 1507}, 042 (2015)
  Erratum: [JHEP {\bf 1611}, 056 (2016)]
  [arXiv:1503.09049 [hep-ex]].
  
\bibitem{Feruglio:2016gvd} 
  F.~Feruglio, P.~Paradisi and A.~Pattori,
  Phys.\ Rev.\ Lett.\  {\bf 118}, no. 1, 011801 (2017)
  [arXiv:1606.00524 [hep-ph]].
  
\bibitem{Feruglio:2017rjo} 
  F.~Feruglio, P.~Paradisi and A.~Pattori,
  arXiv:1705.00929 [hep-ph].

\bibitem{ALEPH:2005ab} 
  S.~Schael {\it et al.} [ALEPH and DELPHI and L3 and OPAL and SLD and LEP Electroweak Working Group and SLD Electroweak Group and SLD Heavy Flavour Group Collaborations],
  Phys.\ Rept.\  {\bf 427}, 257 (2006)
  [hep-ex/0509008].

\bibitem{Kao:2009fg} 
  Y.~Kao and T.~Takeuchi,
  arXiv:0910.4980 [hep-ph].

\bibitem{Barbier:2004ez} 
  R.~Barbier {\it et al.},
  Phys.\ Rept.\  {\bf 420}, 1 (2005)
  [hep-ph/0406039].

\bibitem{Chakraborty:2015bsk} 
  A.~Chakraborty and S.~Chakraborty,
  Phys.\ Rev.\ D {\bf 93}, no. 7, 075035 (2016)
  [arXiv:1511.08874 [hep-ph]].
  
\bibitem{Ibanez:1991hv} 
  L.~E.~Ibanez and G.~G.~Ross,
  Phys.\ Lett.\ B {\bf 260}, 291 (1991).

\bibitem{Smirnov:1996bg} 
  A.~Y.~Smirnov and F.~Vissani,
  Phys.\ Lett.\ B {\bf 380}, 317 (1996)
  [hep-ph/9601387].
  
\bibitem{Nath:2006ut} 
  P.~Nath and P.~Fileviez Perez,
  Phys.\ Rept.\  {\bf 441}, 191 (2007)
  [hep-ph/0601023].

\bibitem{Aaij:2014ora} 
  R.~Aaij {\it et al.} [LHCb Collaboration],
  Phys.\ Rev.\ Lett.\  {\bf 113}, 151601 (2014)
  [arXiv:1406.6482 [hep-ex]].

\bibitem{Aaij:2017vbb} 
  R.~Aaij {\it et al.} [LHCb Collaboration],
  arXiv:1705.05802 [hep-ex].

\bibitem{Bordone:2016gaq} 
  M.~Bordone, G.~Isidori and A.~Pattori,
  Eur.\ Phys.\ J.\ C {\bf 76}, no. 8, 440 (2016)
  [arXiv:1605.07633 [hep-ph]].
  
\bibitem{Hiller:2014ula} 
  G.~Hiller and M.~Schmaltz,
  JHEP {\bf 1502}, 055 (2015)
  [arXiv:1411.4773 [hep-ph]].

\bibitem{Becirevic:2016oho} 
  D.~Becirevic, N.~Kosnik, O.~Sumensari and R.~Zukanovich Funchal,
  JHEP {\bf 1611}, 035 (2016)
  [arXiv:1608.07583 [hep-ph]].
    
\end{thebibliography}
\end{document}